# Fast Deterministic Selection

Andrei Alexandrescu

The D Language Foundation
andrei@erdani.com

**Abstract**

The Median of Medians (also known as BFPRT) algorithm, although a landmark theoretical achievement, is seldom used in practice because it and its variants are slower than simple approaches based on sampling. The main contribution of this paper is a fast linear-time deterministic selection algorithm QUICKSELECTADAPTIVE based on a refined definition of MEDIANOFMEDIANS. The algorithm's performance brings deterministic selection—along with its desirable properties of reproducible runs, predictable run times, and immunity to pathological inputs—in the range of practicality. We demonstrate results on independent and identically distributed random inputs and on normally-distributed inputs. Measurements show that QUICKSELECTADAPTIVE is faster than state-of-the-art baselines.

***Categories and Subject Descriptors*** I.1.2 [*Symbolic and Algebraic Manipulation*]: Algorithms

## 1. Introduction

The selection problem is widely researched and has many applications. In its simplest formulation, selection is finding the $k$th smallest element (also known as the $k$th order statistic) of an array: given an unstructured array $A$, a non-strict order $\leq$ over elements of $A$, and an index $k$, the task is to find the element that would be at position $A[k]$ if the array were fully sorted. A variant that is the focus of this paper is *partition-based selection*: in addition to finding the $k$th smallest element, the algorithm must also permute elements of $A$ such that $A[i] \leq A[k]\ \forall i, 0 \leq i < k$, and $A[k] \leq A[i]\ \forall i, k \leq i < n$. Sorting $A$ solves the problem in $O(n \log n)$, so the challenge is to achieve selection in $O(n)$ time.

Selection often occurs as a step in more complex algorithms such as shortest path, nearest neighbor, and ranking. QUICKSELECT, invented by Hoare in 1961 [15], is the selection algorithm most used in practice [8, 27, 31]. It can be thought of as a partial QUICKSORT [16]: Like QUICKSORT, QUICKSELECT relies on a separate routine to divide the input array into elements less than or equal to and greater than or equal to a chosen element called the pivot. Unlike QUICKSORT which recurses on both subarrays left and right of the pivot, QUICKSELECT only recurses on the side known to contain the $k$th smallest element. (The side to choose is known because the partitioning computes the rank of the pivot.)

The pivot choosing strategy strongly influences QUICKSELECT's performance. If each partitioning step reduces the array portion containing the $k$th order statistic by at least a fixed fraction, QUICKSELECT runs in $O(n)$ time; otherwise, QUICKSELECT is liable to run in quadratic time. Therefore, strategies for pivot picking are a central theme in QUICKSELECT. Some of the most popular and well-studied heuristics choose the median out of a small (1–9) number of elements (either from predetermined positions or sampled uniformly at random) and use that as a pivot [3, 8, 14, 32]. However, such heuristics cannot provide strong worst-case guarantees.

Selection in guaranteed linear time remained an open problem until 1973, when Blum, Floyd, Pratt, Rivest, and Tarjan introduced the seminal "Median of Medians" algorithm [4] (also known as BFPRT after the initials of its authors), an ingenious pivot selection method that works in conjunction with QUICKSELECT. § 3 discusses both QUICKSELECT and MEDIANOFMEDIANS in detail.

Although widely described and studied, MEDIANOFMEDIANS is seldom used in practice. This is because its pivot finding procedure has run time proportional to the input size and is relatively intensive in both element comparisons and swaps. In contrast, sampling-based pivot picking only does a small constant amount of work. In practice, the better quality of the pivot found by MEDIANOFMEDIANS does not justify its higher cost. Therefore, the state of the art in selection with partitioning has been QUICKSELECT in conjunction with simple pivot choosing heuristics.

It would seem that heuristics-based selection has won the server room, leaving deterministic selection to the classroom. However, fast deterministic selection remains desirable for several practical reasons:

- *Immunity to pathological inputs:* Deterministic sampling heuristics (such as "median of first, middle, and last element") are all susceptible to pathological inputs that make QUICKSELECT's run time quadratic. These inputs are easy to generate and contain patterns that are plausible in real data [26, 27, 34]. In order to provide linearity guarantees, many current implementations of QUICKSELECT include code that detects and avoids such situations, at the cost of sometimes falling back to a slower (albeit not quadratic) algorithm, such as MEDIANOFMEDIANS itself. The archetypal example is INTROSELECT [27]. A linear deterministic algorithm that is also fast would avoid these inefficiencies and complications by design.

- *Deterministic behavior:* Randomized pivot selection leaves the input array in a different configuration after each call (even for identical inputs), which makes it difficult to reproduce runs (e.g. for debugging purposes). In contrast, deterministic selection always permutes elements of a given array the same way.

- *Predictability of running time:* In real-time systems (e.g. that compute the median of streaming packets) a randomized pivot choice may cause problems. The first 1–3 pivot choices have a high impact on the overall run time of quickselect with randomized pivots. This makes for a large variance in individual run times [9, 18], even against the same input. Deterministic algorithms have a more predictable run time.

We seek to improve the state of the art in fast deterministic selection, specifically aiming practical large-scale applicability. First, we propose a refined definition of MEDIANOFMEDIANS. The array permutations necessary for finding the pivot also contribute to the partitioning process, thus reducing both overall comparisons and data swaps. We apply this idea to Chen and Dumitrescu's recent REPEATEDSTEP algorithm [6], a MEDIANOFMEDIANS variant particularly amenable to our improvements, and we obtain a competitive baseline.

Second, we add *adaptation* to MEDIANOFMEDIANS. The basic observation is that MEDIANOFMEDIANS is specialized in finding the median, not some arbitrary order statistic. That makes the performance of MEDIANOFMEDIANS degrade considerably when selecting order statistics away from the median (e.g. 25th percentile). We devise simple and efficient specialized algorithms for searching for indexes away from the middle of the searched array. The best algorithm to use is chosen dynamically.

The resulting composite algorithm QUICKSELECTADAPTIVE was measured to be faster than relevant state-of-the-art baselines, which makes it a candidate for solving the selection problem in practical libraries and applications. We propose a generic open-source implementation of QUICKSELECTADAPTIVE for inclusion in the D language's standard library, where it can serve as a reference and basis for porting to other languages and systems.

In the following sections we use the customary pseudocode and algebraic notations to define algorithms. Divisions of integrals are integral with truncation as in many programming languages, e.g. $n/k$ is $\lfloor \frac{n}{k} \rfloor$. When real division is needed, we write $\mathrm{real}(n)/\mathrm{real}(k)$ to emphasize conversion of operands to real numbers prior to the division. Arrays are zero-based so as to avoid minute differences between the pseudocode algorithms and their implementations available online. The length of array $A$ is written as $|A|$. We denote with $A[a:b]$ (if $a < b$) a "slice" of $A$ starting with $A[a]$ and ending with (and including) $A[b-1]$. The slice $A[a:b]$ is empty if $a = b$. Elements of the array are not copied—the slice is a bounded view into the existing array.

The next section discusses related work. § 3 reviews the algorithm definitions we start from. § 4 introduces the improvements we propose to MEDIANOFMEDIANS and REPEATEDSTEP, which are key to better performance. § 5 discusses adaptation for MEDIANOFMEDIANS and derivatives. § 6 includes a few engineering considerations. § 7 presents experiments and results with the proposed algorithms. A summary concludes the paper.

## 2. Related Work

The selection problem has a long history, starting with Lewis Carroll's 1883 essay on the fairness of tennis tournaments (as recounted by Knuth [20, Vol. 3]).

Hoare [15] created the QUICKSELECT algorithm in 1961, which still is the preferred choice of practical implementations, in conjunction with various pivot choosing strategies. Sophisticated running time analyses exist for quickselect [13, 18, 19, 23, 28]. Martínez, Panario, and Viola [24] analyze the behavior of QUICKSELECT with small data sets and propose stopping QUICKSELECT's recursion early and using sorting as an alternative policy below a cutoff, essentially a simple multi-strategy QUICKSELECT. Same authors [25] propose several adaptive sampling strategies for QUICKSELECT that take into account the index searched.

Blum, Floyd, Pratt, Rivest, and Tarjan created the MEDIANOFMEDIANS algorithm [4]. Subsequent work provided variants and improved on its theoretical bounds [5, 11, 12, 21, 36, 37]. Chen and Dumitrescu [6] propose variants of MEDIANOFMEDIANS that group 3 or 4 elements at a time (the original uses groups of 5 or more elements). Most variants offer better bounds and performance than the original, but to date neither has been shown to be competitive against heuristics-based algorithms.

Battiato et al. [2] describe Sicilian Median Selection, a fast algorithm for computing an approximate median. It could be considered the transitive closure of Chen and Dumitrescu's REPEATEDSTEP algorithm. In spite of good average performance, the algorithm's worst-case pivot quality is insufficient for guaranteeing exact selection in linear time.

# 3. Background

## 3.1 QUICKSELECT and MEDIANOFMEDIANS

We provide a quick overview of the QUICKSELECT and MEDIANOFMEDIANS algorithms [8, 20] with an emphasis on practical implementations.

Algorithm 1 illustrates QUICKSELECT. It takes as parameters not only the data (array $A$ and index $k$), but also a partitioning primitive called PARTITION, in a higher-order function fashion. PARTITION$(A, k)$ returns an index $p$ (called pivot) with $0 \leq p < |A|$ and also partitions $A$ such that $A[j] \leq A[p] \; \forall j, 0 \leq j < p$, and $A[p] \leq A[j] \; \forall j, p < j < |A|$. QUICKSELECT, in a divide and conquer manner, uses the pivot to either reduce the portion of the array searched from the left when $p < k$, reduce it from the right when $k < p$, or finish the search when $p = k$. Overall, QUICKSELECT uses an approximate partitioning primitive repeatedly to obtain a precise partition around the specific index $k$.

---

**Algorithm 1:** QUICKSELECT

**Data:** PARTITION, $A$, $k$ with $|A| > 0, 0 \leq k < |A|$
**Result:** Puts $k$th smallest element of $A$ in $A[k]$ and partitions $A$ around it.

1 **while true do**
2     $p \leftarrow$ PARTITION$(A, k)$;
3     **if** $p = k$ **then**
4        **return**;
5     **end**
6     **if** $p > k$ **then**
7        $A \leftarrow A[0 : p]$;
8     **else**
9        $k \leftarrow k - p - 1$;
10       $A \leftarrow A[p + 1 : |A|]$;
11    **end**
12 **end**

---

The running time is linear if PARTITION$(A, k)$ is linear and able to return a pivot $p$ within some fixed fraction $f < 1$ from either end of $A$, i.e. $p > f|A|$ and $p < (1 - f)|A|$, which allows QUICKSELECT to eliminate at least $f|A|$ elements at each pass. In that case, the time complexity $T(n)$ (depending on $n = |A|$) of QUICKSELECT satisfies:

$$T(n) \leq T((1-f)n) + O(n) \quad (1)$$

which results in $T(n) = O(n)$ by the Master Theorem [30].

There is no required relationship between $k$ and $p$, and in fact many partitioning routines do not use $k$. It may be considered a hint. Ideally, the partitioning primitive would return a pivot equal to $k$, or close to $k$ (greater than it if $k < \frac{|A|}{2}$, smaller otherwise). That allows QUICKSELECT to either end the search or eliminate many elements at each pass. Contemporary approaches prevalently ignore $k$ and attempt to partition close to the median, strategy shown recently to be suboptimal [25].

Most pivot selection schemes are based on *heuristics*: they choose an array element likely to be not too close to the minimum or maximum, and then use HOAREPARTITION [16] (Algorithm 2), which partitions the array around that element in linear time. Some popular and well researched heuristics include:

- *Random pivot [8]*: the pivot is chosen uniformly at random from $A$;
- *Median of 3 [32]*: the pivot is the median of $A[0]$, $A[|A|/2]$, and $A[|A| - 1]$;
- *Median of 3 randomized [17, 32]*: the pivot is the median of three elements chosen uniformly at random from $A$;
- *Tukey's ninther [3, 33]*: Choose $A[0]$, $A[|A|/8]$, $A[2|A|/8]$, $A[3|A|/8]$, $A[4|A|/8]$, $A[5|A|/8]$, $A[6|A|/8]$, $A[7|A|/8]$, and $A[|A| - 1]$, then compute the median of first three, middle three, and last three, and finally the pivot is the median of the three medians;
- *Tukey's ninther randomized [34]*: Similar to the ninther, but the nine samples are chosen uniformly at random from $A$.

---

**Algorithm 2:** HOAREPARTITION

**Data:** $A$, $p$ with $|A| > 0, 0 \leq p < |A|$
**Result:** $p'$, the new position of $A[p]$; $A$ partitioned at $A[p']$

1 SWAP$(A[p], A[0])$;
2 $a = 1$;   $b = |A| - 1$;
3 loop: **while true do**
4     **while true do**
5        **if** $a > b$ **then break** loop;
6        **if** $A[a] \geq A[0]$ **then break**;
7        $a \leftarrow a + 1$;
8     **end**
9     **while** $A[0] < A[b]$ **do** $b \leftarrow b - 1$;
10    **if** $a \geq b$ **then break**;
11    SWAP$(A[a], A[b])$;
12    $a \leftarrow a + 1$;
13    $b \leftarrow b - 1$;
14 **end**
15 SWAP$(A[0], A[a - 1])$;
16 **return** $a - 1$;

---

No heuristics-based approach can provide a good worst-case run time guarantee. However, the constant-time speed of computing the pivot compensates on average for its poor quality.

In contrast, MEDIANOFMEDIANS spends more time to systematically guarantee good pivot choices. Algorithm 3 illustrates the prevalent implementation of MEDIANOFMEDIANS [7, 10, 29, 35].

**Algorithm 3:** BFPRTBASELINE

   **Data:** $A, k$ ($k$ not used)
   **Result:** Pivot $0 \leq p < |A|$; $A$ partitioned at $A[p]$
1  **if** $|A| < 5$ **then**
2     |  **return** HOAREPARTITION$(A, |A|/2)$;
3  **end**
4  $i \leftarrow 0;\ j \leftarrow 0$;
5  **while** $i + 4 < |A|$ **do**
6     |  MEDIAN5$(A, i, i+1, i+2, i+3, i+4)$;
7     |  SWAP$(A[i+2], A[j])$;    // median to 1st quintile
8     |  $i \leftarrow i + 5$;
9     |  $j \leftarrow j + 1$;
10 **end**
11 QUICKSELECT(BFPRTBASELINE, $A[0:j], j/2$);
12 **return** HOAREPARTITION$(A, j/2)$;

**Algorithm 4:** MEDIAN5

   **Data:** $A, a, b, c, d, e$
   **Result:** Puts median of $A[a], A[b], A[c], A[d], A[e]$ in
                      $A[c]$ and leaves the smaller items in $A[a], A[b]$
                        and the larger ones in $A[d], A[e]$.
1  **if** $A[c] < A[a]$ **then** SWAP$(A[a], A[c])$;
2  **if** $A[d] < A[b]$ **then** SWAP$(A[b], A[d])$;
3  **if** $A[d] < A[c]$ **then**
4     |  SWAP$(A[c], A[d])$;
5     |  SWAP$(A[a], A[b])$;
6  **end**
7  **if** $A[e] < A[b]$ **then** SWAP$(A[b], A[e])$;
8  **if** $A[e] < A[c]$ **then**
9     |  SWAP$(A[c], A[e])$;
10    |  **if** $A[c] < A[a]$ **then** SWAP$(A[a], A[c])$;
11 **else**
12    |  **if** $A[c] < A[b]$ **then** SWAP$(A[b], A[c])$;
13 **end**

The algorithm first computes medians of groups of 5 elements at a time, for a total of $\frac{|A|}{5}$ groups. The primitive MEDIAN5 takes an array and five distinct indexes, computes their median with a small specialized algorithm, and places it in the third index received. (The last $|A| \bmod 5$ elements of the array are ignored for simplicity.) Computing (recursively) the median of these medians yields a pivot with a desirable property. There are $\frac{|A|}{10}$ elements less than or equal to the pivot, but each of those is the median of some 5 distinct elements, so for each of those $\frac{|A|}{10}$ elements there are 2 others also less than or equal to the pivot. That guarantees at least $\frac{3|A|}{10}$ elements less than or equal to the pivot. By symmetry, at least $\frac{3|A|}{10}$ elements in the array are greater than or equal to the pivot.

Selection is initiated by invoking QUICKSELECT(BFPRTBASELINE, $A, k$). Line 11 passes the name of the algorithm (BFPRTBASELINE) as an argument to QUICKSELECT. This makes the two algorithms mutually recursive: BFPRTBASELINE is the partitioning primitive used by QUICKSELECT, and in turn uses QUICKSELECT itself.

MEDIAN5 is used intensively by the algorithm and therefore deserves a careful implementation. A good balance needs to be found between comparisons, swaps, and function size. We chose a function with 6 comparisons and 0–7 swaps (only 3.13 on average against distinct random numbers), shown in Algorithm 4. (To simplify notation we use $A[a] < A[b]$ as a shortcut for $\neg(A[b] \leq A[a])$.) The function is also idempotent: if the five slots are already partitioned, it performs no swaps.

To prove linearity for MEDIANOFMEDIANS, let us look at the number of comparisons $C(n)$ that BFPRTBASELINE executes as a function of $n = |A|$. In the worst case, we have:

$$C(n) \leq C\left(\frac{n}{5}\right) + C\left(\frac{7n}{10}\right) + \frac{6n}{5} + n \qquad (2)$$

where the first term accounts for the recursive call to compute the median of medians, the second is the recursive call in QUICKSELECT (in the worst case $\frac{3n}{10}$ elements are eliminated in one partitioning step so $\frac{7n}{10}$ are left), the third is the cost of computing the medians of five, and the last is the cost of HOAREPARTITION. Consequently $C(n) \leq 22n$.

The number of swaps $S(n)$ is also of interest. That satisfies the recurrence:

$$S(n) \leq S\left(\frac{n}{5}\right) + S\left(\frac{7n}{10}\right) + \frac{7n}{5} + \frac{n - \frac{n}{10}}{2} \qquad (3)$$

The terms correspond to those for $C(n)$. Our median of five routine uses at most 7 swaps. Also in the last term we eliminate the first $\frac{n}{10}$ elements of the array. Those are known to be no greater than the median so they will not be swapped. Consequently $S(n) \leq \frac{37n}{2}$.

Although Algorithm 3 uses groups of 5, any greater constant works as well (for the implementation and the linearity proof above); in fact, Blum et al. discuss group sizes of 15 and 21 elements. Group size 5 is prevalently used today because it is the most effective in practice.

### 3.2 The REPEATEDSTEP Algorithm

Recently Chen and Dumitrescu [6] proposed MEDIANOF-MEDIANS variants that use groups of 3 or 4 elements yet have linear worst-case running time, disproving long-standing conjectures to the contrary. Algorithm 5 shows the pseudocode of their REPEATEDSTEP algorithm with a group size of 3.

**Algorithm 5:** REPEATEDSTEP

**Data:** $A$, $k$ ($k$ not used)
**Result:** Pivot $0 \leq p < |A|$; $A$ partitioned at $A[p]$

1 **if** $|A| < 9$ **then**
2     **return** HOAREPARTITION($A, |A|/2$);
3 **end**
4 $i \leftarrow 0$; $j \leftarrow 0$;
5 **while** $i + 2 < |A|$ **do**
6     MEDIAN3($A, i, i+1, i+2$);
7     SWAP($A[i+1], A[j]$);        // median to 1st tertile
8     $i \leftarrow i + 3$;
9     $j \leftarrow j + 1$;
10 **end**
11 $i \leftarrow 0$; $m \leftarrow 0$;
12 **while** $i + 2 < j$ **do**
13     MEDIAN3($A, i, i+1, i+2$);
14     SWAP($A[i+1], A[m]$);        // median to 1st 9th-ile
15     $i \leftarrow i + 3$;
16     $m \leftarrow m + 1$;
17 **end**
18 QUICKSELECT(REPEATEDSTEP, $A[0:m], m/2$);
19 **return** HOAREPARTITION($A, m/2$);

The algorithm is similar to BFPRTBASELINE with group size changed to 3, with one important addition: the median of medians step is repeated, essentially choosing the pivot as the median of medians of medians of three (sic), instead of median of medians of three. This degrades the quality of the pivot because of the double approximation: the pivot is guaranteed to be within $\frac{2n}{9}$ elements from either edge of the array (weaker than $\frac{n}{3}$ without the repeated step, or $\frac{3n}{10}$ if using groups of 5). However, this loss in quality comes with an improvement in speed: the median of medians computation only needs to recurse on $\frac{n}{9}$ elements instead of $\frac{n}{3}$.

Intuitively, trading off some pivot quality for faster processing in MEDIANOFMEDIANS is a good idea, since pivot heuristics that provide relatively poor pivot estimates very fast perform so well in practice. There are advantages specific to using groups of 3—computing the median of 3 is simpler and takes 2–3 comparisons whereas the median of 5 requires 6 comparisons. So it is expected that REPEATEDSTEP improves performance over BFPRTBASELINE. To assess that, let us estimate the number of comparisons $C(n)$ performed by REPEATEDSTEP from the recurrence:

$$C(n) \leq C\left(\frac{n}{9}\right) + C\left(\frac{7n}{9}\right) + \frac{3n}{3} + \frac{3n}{9} + n \quad (4)$$

where the first term is the cost of recursion for computing the median of medians of medians, the second is the worst-case time spent processing the remaining elements, and the last three terms account respectively for computing the medians of three (up to three comparisons per group of three), computing the medians of medians of three, and the partitioning. Consequently $C(n) \leq 21n$.

We now calculate a bound for $S(n)$. Each median of three uses up to two swaps, to which we add one for placing the median at the front of the array. The partitioning costs at most $\frac{n}{2}$ swaps in general, but in this case we already have the first $\frac{n}{18}$ elements of the array known to be no greater than the pivot, so they will not be swapped. We obtain:

$$S(n) \leq S\left(\frac{n}{9}\right) + S\left(\frac{7n}{9}\right) + n + \frac{n}{3} + \frac{n - \frac{n}{18}}{2} \quad (5)$$

which solves to $S(n) \leq \frac{65n}{4}$. Both bounds are better than those of BFPRTBASELINE.

## 4. Improving Layout

We now start improving these algorithms, first by laying out data in a more efficient manner. Note that BFPRTBASELINE (Algorithm 3) already embeds an important optimization—it reuses the first quintile of $A$ for storing the medians. This approach avoids the additional overhead of creating and using a separate temporary array to store the medians of 5 and is in keeping with today's most frequently encountered implementations [7, 10, 29, 35].

One key insight motivates a different layout choice: we aim to make the comparisons and swaps performed during the medians of 5 computation also count toward the partitioning. MEDIANOFMEDIANS organizes the array in groups of 5 and computes the median of each. In addition, in each group of 5, the 2 smaller elements are to the left, and the 2 larger ones are to the right of the median. That imparts quite a non-trivial implicit structure onto the array in addition to computing the pivot. However, that structure is of no use to HOAREPARTITION. Ideally, that structure should be embedded in the array in a form favorable to the subsequent partitioning step; conversely, the partitioning step should make maximal use of the implicit information established in the array by the medians of 5 stage. Indeed, the original MEDIANOFMEDIANS paper [4] provides an optimized algorithm called PICK1 that makes use of the implicit structuring of the input, but is still not efficient in practice.

Our approach (BFPRTIMPROVED shown in Algorithm 6) is to make the groups of 5 *non-contiguous* and lay them out in a manner that is advantageous for the partitioning step. We place the subarray of medians in the third quintile, the smaller values to its left, and the larger values to its right. This is efficient because the third quintile is in the middle of the entire array. After that quintile is partitioned around its own median (by the recursive call to QUICKSELECT), it already is globally partitioned properly around the pivot; there is no more need to visit it. That way, the medians of 5 computation step contributes one fifth of the final result at no additional cost. The subsequent partitioning step only

5        2016/7/27

needs to partition quintiles 1, 2, 4, and 5. Better yet, quintiles 1 and 2 contain the statistically smaller elements and quintiles 4 and 5 contain the statistically larger elements so the medians of 5 computation also saves on swapping during partitioning.

In order to distribute the groups of 5 appropriately, we take advantage of the fact that the MEDIAN5 primitive (Algorithm 4) accepts any five distinct indexes, not only consecutive ones. So we can choose any convenient grouping and iteration schedule. Let $f = \lfloor \frac{|A|}{5} \rfloor$ such that quintiles start at $A[0]$, $A[f]$, $A[2f]$, $A[3f]$, and $A[4f]$. Then we choose the first group as MEDIAN5$(A, 0, 1, 2f, 3f, 3f+1)$. These are respectively the leftmost two elements to the left of the third quintile, the first element of the third quintile, and the leftmost two elements to the right of the third quintile. Iteration proceeds with median calls following the schedule MEDIAN5$(A, i, i+1, 2f+j, 3f+i, 3f+i+1)$, with $i$ iterating with step 2 and $j$ iterating with step 1. The entire array is covered when $j = f$.

Computing medians of five using the iteration strategy above and then computing the median of the third quintile leaves the array with the following layout:

- The elements likely less than or equal to the pivot (at least $\frac{3n}{10}$) are already to the left of $A$'s middle.
- The elements likely larger than or equal to the pivot (again at least $\frac{3n}{10}$) are already to the right of $A$'s middle.
- The medians of five are in the third quintile of the array and are already partitioned around the pivot, so there is no more need for the partitioning step to compare $\frac{n}{5}$ elements.

This leaves the array well suited for partitioning with only light swapping. The subarrays $A[0 : 2f]$ and $A[3f : |A|]$ need to be visited and partitioned properly. This work is carried by EXPANDPARTITION (not shown as it is trivial yet replete with detail), a slightly modified HOAREPARTITION that takes into account the already-partitioned subarray around the pivot. The call EXPANDPARTITION$(A, a, p, b)$ proceeds by the following scheme, starting with $i = 0$ and $j = |A| - 1$ and moving them toward $a$ and $b$, respectively:

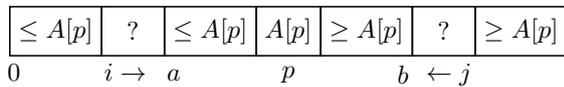

The procedure swaps as many elements as possible between $A[i : a]$ and $A[b+1 : j+1]$, because that is the most efficient use of swapping—one swap puts two elements in their final slots. Often there will be some asymmetry (one of $i$ and $j$ reaches its limit before the other) so the pivot position will need to shift to the left or to the right. EXPANDPARTITION returns the final position of the pivot, which BFPRTIMPROVED forwards to the caller. EXPANDPARTITION$(A, a, b)$ performs $\max(a, |A| - b)$ comparisons and at most $\max(a, |A| - b)$ swaps.

**Algorithm 6:** BFPRTIMPROVED

**Data:** $A$, $k$ ($k$ not used)
**Result:** Pivot $p$, $0 \leq p < |A|$

1 **if** $|A| < 5$ **then**
2     **return** HOAREPARTITION$(A, |A|/2)$;
3 **end**
4 $f \leftarrow |A|/5$; $i \leftarrow 0$;
5 **for** $j \leftarrow 2f$ **through** $3f - 1$ **do**
6     MEDIAN5$(A, i, i+1, j, 3f+i, 3f+i+1)$;
7     $i \leftarrow i + 2$;
8 **end**
9 QUICKSELECT(BFPRTIMPROVED, $A[2f : 3f]$, $f/2$);
   **return** EXPANDPARTITION$(A, 2f, 2f + f/2, 3f - 1)$;

We now derive worst-case bounds for BFPRTIMPROVED depending on $n = |A|$. The only difference in computing $C(n)$ is one fifth of the elements do not need to be compared any more, which changes the last term:

$$C(n) \leq C\left(\frac{n}{5}\right) + C\left(\frac{7n}{10}\right) + \frac{6n}{5} + \frac{4n}{5} \quad (6)$$

which leads to $C(n) \leq 20n$. For swaps:

$$S(n) \leq S\left(\frac{n}{5}\right) + S\left(\frac{7n}{10}\right) + \frac{7n}{5} + \frac{2n}{10} \quad (7)$$

The last term changed because there are at most $\frac{2n}{10}$ elements on either side to swap across the pivot, and the algorithm puts two elements in the right place with one swap whenever possible. The equation results in $S(n) \leq 16n$. By these estimates, BFPRTIMPROVED reduces comparisons by about 9% and swaps by about 14% over BFPRTBASELINE.

### 4.1 Improving Layout for REPEATEDSTEP

Let us now apply the same layout idea to the REPEATEDSTEP algorithm. The result is shown in Algorithm 7. The loop starting at line 5 computes medians of 3 noncontiguous elements using MEDIAN3 (trivial, not shown): first from the first tertile, second from the middle tertile, and third from the last tertile. The three indexes move in lockstep, so only one iteration variable is needed. Similarly, the next loop uses the same iteration schedule to compute medians of three for the middle tertile. This leaves the medians of medians of medians of 3 in the fifth 9th ile (middle "novile") of the array.

The call to QUICKSELECT partitions the mid 9th ile around its (exact) median. Right after line 11 the following conditions are satisfied:

- Pivot value is in $A[p]$;
- $A[4f : p]$ contains only elements less than or equal to the pivot;

**Algorithm 7:** REPEATEDSTEPIMPROVED

    **Data:** $A$, $k$ ($k$ not used)
    **Result:** Pivot $p$, $0 \leq p < |A|$; $A$ partitioned at $p$
1  **if** $|A| < 9$ **then**
2     **return** HOAREPARTITION($A, |A|/2$);
3  **end**
4  $f \leftarrow |A|/9$;
5  **for** $i \leftarrow 3f$ **through** $6f - 1$ **do**
6     MEDIAN3($A, i - 3f, i, i + 3f$);
7  **end**
8  **for** $i \leftarrow 4f$ **through** $5f - 1$ **do**
9     MEDIAN3($A, i - f, i, i + f$);
10 **end**
11 QUICKSELECT(REPEATEDSTEPIMPROVED, $A[4f : 5f], f/2$);
12 **return** EXPANDPARTITION($A, 4f, 4f + f/2, 5f - 1$);

- $A[p+1 : 5f]$ contains only elements greater than or equal to the pivot;
- At least $\frac{2n}{9}$ elements in $A[0 : 4f]$ are less than or equal to the pivot;
- At least $\frac{2n}{9}$ elements in $A[5f : |A|]$ are greater than or equal to the pivot.

Let us derive bounds for $C(n)$ and $S(n)$ for this algorithm. The number of comparisons in REPEATEDSTEPIMPROVED satisfies the recurrence:

$$C(n) \leq C\left(\frac{n}{9}\right) + C\left(\frac{7n}{9}\right) + n + \frac{n}{3} + \frac{8n}{9} \qquad (8)$$

The only difference from the basic algorithm is the last term, which is slightly smaller in this case because we don't revisit the middle 9$^{\text{th}}$ile during partitioning. This solves to $C(n) \leq 20n$ (5% better than REPEATEDSTEP).

Each MEDIAN3 call performs at most two swaps. Due to the layout choice the partitioning step already leaves at least $\frac{2n}{9}$ elements less than or equal to the pivot to the pivot's left, and at least as many elements greater than or equal to the pivot to its right, so in the worst case the algorithm needs to swap $\frac{5n}{18}$ elements from the left side and $\frac{5n}{18}$ elements from the right. However, the swaps don't sum from the two sides because one swap operation takes care of two elements wherever possible. So the total number of swaps performed by EXPANDPARTITION is at most $\frac{5n}{18}$.

$$S(n) \leq S\left(\frac{n}{9}\right) + S\left(\frac{7n}{9}\right) + \frac{2n}{3} + \frac{2n}{9} + \frac{5n}{18} \qquad (9)$$

Consequently $S(n) \leq \frac{21n}{2}$, over 35% fewer than REPEATEDSTEP algorithm.

These bounds are adequate for proving linearity and useful as proxies, but not necessarily tight (producing worst-case inputs for MEDIANOFMEDIANS remains an open problem); the measured speed improvements are larger. Although algorithms with better upper bounds (both comparisons and swaps) for selection have been devised [5, 21], they are more complex and entail large hidden costs; to date neither of those has been shown to be efficient in practice.

This layout change has been crucial for obtaining efficiency from MEDIANOFMEDIANS. The improved layout not only drastically improves performance, but also allows further optimizations to build upon it. REPEATEDSTEPIMPROVED is our core algorithm for fast deterministic selection.

## 5. Adaptation

The performance of REPEATEDSTEPIMPROVED compares favorably with state-of-the-art baselines when searching the median. However, performance degrades quickly as the index sought migrates from the middle of the array toward either of its ends. This is because REPEATEDSTEPIMPROVED (like all MEDIANOFMEDIANS variations discussed so far) is specialized in finding a good approximate *median*, not a good approximate $k$th order statistic for any $k$. We acknowledged that by mentioning "$k$ not used" in each algorithm's parameters section.

However, $k$ may be very informative especially when far from the middle. For an extreme example, consider searching a position $k$ only a few slots from one edge of the array. Then any heuristic will produce a good-quality pivot very quickly simply because the range of good pivots is large. In contrast, a MEDIANOFMEDIANS algorithm would do considerable work to eliminate about half the array. This continues through the end of the search—heuristics eliminate swaths of the array at constant cost, whereas MEDIANOFMEDIANS pedantically reduces the array by about half at every pass at linear cost. The average time complexity is linear for either approach, but heuristics-based algorithms will complete faster.

Our approach to introducing adaptation is two-pronged. First, we use simple interpolation to improve the speed of REPEATEDSTEPIMPROVED with off-median indexes. Second, we define specialized algorithms for indexes close to one edge of the array. In the following we focus on small indexes (i.e. less than $\frac{|A|}{2}$). This is without loss of generality; large indexes work symmetrically.

### 5.1 Making REPEATEDSTEPIMPROVED Adaptive

Line 11 of REPEATEDSTEPIMPROVED computes the median of the fourth 9$^{\text{th}}$ile with the call QUICKSELECT(REPEATEDSTEPIMPROVED, $A[4f : 5f], f/2$). The last argument divides the subarray in half, computing an approximate median of the entire array. However, if $k \neq \frac{|A|}{2}$, we are more interested in a pivot close to $k$ than an approximate median.

To make REPEATEDSTEPIMPROVED adaptive, we adopt a simple interpolation approach: replace $f/2$ with $kf/|A|$ in the call above, idea first proposed and analyzed by Martínez et al. [25] with the name "proportional-of-s". That way, $k$ is proportionally represented in the fraction by which the subarray is partitioned. There is no change if $k = |A|/2$; as $k$ gets smaller the pivot computed will move left, and as $k$ gets larger the pivot will move right. Algorithm 8 shows the pseudocode of the resulting algorithm REPEATEDSTEPADAPTIVE, which is the first discussed here to use $k$.

---

**Algorithm 8:** REPEATEDSTEPADAPTIVE

**Data:** $A, k$
**Result:** Pivot $p$, $0 \leq p < |A|$; $A$ partitioned at $A[p]$

1 **if** $|A| < 9$ **then**
2    **return** HOAREPARTITION$(A, |A|/2)$;
3 **end**
4 $f \leftarrow |A|/9$;
5 **for** $i \leftarrow 3f$ **through** $6f - 1$ **do**
6    MEDIAN3$(A, i - 3f, i, i + 3f)$;
7 **end**
8 **for** $i \leftarrow 4f$ **through** $5f - 1$ **do**
9    MEDIAN3$(A, i - f, i, i + f)$;
10 **end**
11 QUICKSELECT(REPEATEDSTEPADAPTIVE, $A[4f : 5f], kf/|A|$);
12 **return** EXPANDPARTITION$(A, 4f, 4f + kf/|A|, 5f - 1)$;

---

## 5.2 Searching for Indexes Left of Median

Performance of REPEATEDSTEPADAPTIVE is better than REPEATEDSTEPIMPROVED when $k$ is off of $\frac{|A|}{2}$. However, as $k$ gets further away from the middle, the number of swaps performed by REPEATEDSTEPADAPTIVE increases. This is because the 9$^{\text{th}}$ile used for recursion is still in the middle of the array, so it is suboptimally placed; EXPANDPARTITION needs to shift elements from that middle 9$^{\text{th}}$ile toward the pivot, which is off the middle.

When $k$ is far away from the median, a better way to improve performance is to partition the array *asymmetrically*. If the index sought is to the left of the median, placing the subarray of medians on the left and ensuring a larger margin on the right is better. That way the search has less swapping to do and is more likely to eliminate a large subarray. To get asymmetric margins, we modify the group size and placement as follows.

Let us start by noting that REPEATEDSTEPADAPTIVE does not necessarily need to use groups of 3. It may use different group sizes for the first and the repeated step, as long as linearity can be proven. An even group size is of particular interest to us because it partitions the array asymmetrically. Consider taking the lower median of 4 in the first step, then the median of 3 in the second step and use that as the pivot. We do so with the help of the trivial routine LOWERMEDIAN4$(A, a, b, c, d)$ (not shown), which places the lower median of $A[a], A[b], A[c], A[d]$ in $A[b]$ and the minimum in $A[a]$. In that case, after the two partition steps (with group sizes 4 then 3) are finished, there are at least $\frac{|A|}{6}$ elements less than or equal to the pivot on the left, and $\frac{|A|}{4}$ elements greater than or equal to the pivot on the right. (If the upper median of 4 were used, the numbers would be reversed.) The partitioned subarray is asymmetrically placed at $A[|A|/3 : 5|A|/12]$, which reduces swaps if the pivot is also on the left.

Algorithm 9 defines REPEATEDSTEPLEFT. The call to QUICKSELECT uses the same interpolation approach as REPEATEDSTEPADAPTIVE.

---

**Algorithm 9:** REPEATEDSTEPLEFT

**Data:** $A, k$
**Result:** Pivot $p$, $0 \leq p < |A|$; $A$ partitioned at $A[p]$

1 **if** $|A| < 12$ **then**
2    **return** HOAREPARTITION$(A, |A|/2)$;
3 **end**
4 $f \leftarrow |A|/4$;
5 **for** $i \leftarrow 0$ **through** $f - 1$ **do**
6    LOWERMEDIAN4$(A, i, i + f, i + 2f, i + 3f)$;
7 **end**
8 $f' \leftarrow f/3$;
9 **for** $i \leftarrow f$ **through** $f + f' - 1$ **do**
10    MEDIAN3$(A, i, i + f', i + 2f')$;
11 **end**
12 QUICKSELECT(REPEATEDSTEPLEFT, $A[f : f + f'], kf'/|A|$);
13 **return** EXPANDPARTITION$(A, f, f + kf'/|A|, f + f' - 1)$;

---

The number of comparisons in the worst case obeys:

$$C(n) \leq C\left(\frac{n}{12}\right) + C\left(\frac{5n}{6}\right) + \frac{4n}{4} + \frac{3n}{12} + n \quad (10)$$

with the same term positions as in Eq. 4 and the note that lower/upper median of 4 requires at most 4 comparisons. For this asymmetric algorithm, $C(n)$ is also linear and bounded by $21n$, even if inefficiently used for all values of $k$.

REPEATEDSTEPLEFT is best used with indexes $k$ below $A$'s middle (by a threshold we established experimentally). A similarly-defined routine REPEATEDSTEPRIGHT is to be used with indexes above the middle of $A$. For sought indexes close to the median, REPEATEDSTEPADAPTIVE offers best results.

## 5.3 Searching for Indexes Far Left of Median

Consider now searching for very small indexes. In that case the motivation is high to *not* find a pivot to the left of the index, because that would only eliminate very little of the input. Martínez et al. discuss this risk for their related proportional-of-3 search strategy [25]. So we design an algorithm to find $p$ such that $p \geq k$.

A different schema is useful here. REPEATEDSTEPFAR-LEFT (Algorithm 10) takes the lower median of 4 in the first step, like REPEATEDSTEPLEFT. Then, instead of the median of 3 in the second step, it takes the *minimum* of 3.

---

**Algorithm 10:** REPEATEDSTEPFARLEFT

**Data:** $A, k$
**Result:** Pivot $p$, $0 \leq p < |A|$; $A$ partitioned at $A[p]$

1. **if** $|A| < 12$ **then**
2.     **return** HOAREPARTITION$(A, |A|/2)$;
3. **end**
4. $f \leftarrow |A|/4$;
5. **for** $i \leftarrow f$ **through** $2f - 1$ **do**
6.     LOWERMEDIAN4$(A, i - f, i, i + f, i + 2f)$;
7. **end**
8. $f' \leftarrow f/3$;
9. **for** $i \leftarrow f$ **through** $f + f' - 1$ **do**
10.     **if** $A[i + f'] < A[i]$ **then** SWAP$(A[i + f'], A[i])$;
11.     **if** $A[i + 2f'] < A[i]$ **then** SWAP$(A[i + 2f'], A[i])$;
12. **end**
13. QUICKSELECT(REPEATEDSTEPFARLEFT, $A[f : f + f'], kf'/|A|$);
14. **return** EXPANDPARTITION$(A, f, f + kf'/|A|, f + f' - 1)$;

---

In that case, there are at least $\frac{|A|}{12}$ elements less than or equal to the pivot on the left, and $\frac{3|A|}{8}$ elements greater than or equal to the pivot on the right. (If the upper median was chosen, the numbers would be reversed.) The subarray passed down to QUICKSELECT is placed at $A[|A|/4 : |A|/3]$. $C(n)$ obeys:

$$C(n) \leq C\left(\frac{n}{12}\right) + C\left(\frac{5n}{6}\right) + \frac{4n}{4} + \frac{3n}{12} + n \qquad (11)$$

with the same term positions as in Eq. 4. For this algorithm, $C(n)$ is also linear and bounded by $21n$. However, if REPEATEDSTEPFARLEFT is used only when $k \leq \frac{|A|}{12}$, then the pivot is guaranteed on the right so the bound gets better because $\frac{3|A|}{8}$ or more elements are discarded.

The algorithm REPEATEDSTEPFARRIGHT (not shown) is defined symmetrically with REPEATEDSTEPFARLEFT: it takes the upper medians of 4 and places them in the 3rd quartile, then within that quartile it takes the maximum of 3. The margins are $\frac{3|A|}{8}$ on the left and $\frac{|A|}{12}$ on the right, and the subarray passed down to QUICKSELECT is placed at $A[2|A|/3 : 3|A|/4]$.

## 5.4 Choosing Strategy Dynamically in QUICKSELECT

How to tie these specialized algorithms together? We defined QUICKSELECT as a higher-order function parameterized by the partitioning primitive used. That made it convenient to discuss and analyze a variety of partitioning algorithms using the same QUICKSELECT skeleton.

At this point, in order to implement full adaptation, we need to dynamically choose the partitioning algorithm from among REPEATEDSTEPADAPTIVE, REPEATED-STEPLEFT, REPEATEDSTEPRIGHT, REPEATEDSTEPFAR-LEFT, and REPEATEDSTEPFARRIGHT. A good place to decide strategy is the QUICKSELECT routine itself, which has access to all information needed and controls iteration. Before each partitioning step, a partitioning algorithm is chosen depending on the relationship between $|A|$ and $k$. After partitioning, both $A$ and $k$ are modified and a new decision is made, until the search is over. QUICKSELECTADAPTIVE (Algorithm 11) embodies this idea.

The cutoff $\frac{1.0}{12.0}$ is dictated by the lower margin of RE-PEATEDSTEPFARLEFT and REPEATEDSTEPFARRIGHT; as mentioned, in those cases we never want the pivot to be between the sought index and the closest array edge. The cutoff $\frac{7.0}{16.0}$ has been chosen experimentally.

## 6. Engineering Considerations

The deterministic partitioning algorithms presented so far do not "cheat" (by estimating from samples, making assumptions about data distribution etc): they look at the entire data set and provide guaranteed margins. The absolute cost of a search is front-loaded, during the first few iterations of QUICKSELECTADAPTIVE, when the data examined is largest. For large data sets with usual distribution statistics, a good sampling strategy that finds a good pivot without inspecting the entire data may make a large speed difference. It is therefore very tempting for the optimization-minded engineer to do away with rigorous determinism and try a few shots at sampling when the data size is large.

However, we resisted the lure of switching to one of the classic heuristics. When data is large, indeed the reward of a cheap good pivot is excellent, but conversely the cost of a bad pivot choice is high, and all heuristics may compute an arbitrarily bad pivot.

We adopted an engineering solution that offers good speed improvements on average for only moderate slowdown in the worst case and a minor code change. Each search starts optimistically with a "sampling mode" flag on. As long as sampling is enabled, the first median step in all REPEATEDSTEP algorithm variations is skipped. For example, in Algorithm 7, the loop at line 5 is not executed if sampling is on. This makes the pivot to be evaluated only from the mid tertile; the first and last tertile are not inspected at

**Algorithm 11:** QUICKSELECTADAPTIVE

**Data:** $A$, $k$ with $0 \leq k < |A|$
**Result:** Puts $k$th smallest element of $A$ in $A[k]$ and partitions $A$ around it.

1 **while** true **do**
2     $r \leftarrow \text{real}(k)/\text{real}(|A|)$;
3     **if** $|A| < 12$ **then**
4        $p \leftarrow$ HOAREPARTITION$(A, |A|/2)$;
5     **else if** $r \leq 7.0/16.0$ **then**
6        **if** $r \leq 1.0/12.0$ **then**
7           $p \leftarrow$ REPEATEDSTEPFARLEFT$(A, k)$;
8        **else**
9           $p \leftarrow$ REPEATEDSTEPLEFT$(A, k)$;
10        **end**
11     **else if** $r \geq 1.0 - 7.0/16.0$ **then**
12        **if** $r \geq 1.0 - 1.0/12.0$ **then**
13           $p \leftarrow$ REPEATEDSTEPFARRIGHT$(A, k)$;
14        **else**
15           $p \leftarrow$ REPEATEDSTEPRIGHT$(A, k)$;
16        **end**
17     **else**
18        $p \leftarrow$ REPEATEDSTEPIMPROVED$(A, k)$;
19     **end**
20     **if** $p = k$ **then return**;
21     **if** $p > k$ **then**
22        $A \leftarrow A[0 : p]$;
23     **else**
24        $i \leftarrow k - p - 1$;
25        $A \leftarrow A[p + 1 : |A|]$;
26     **end**
27 **end**

all, so this mechanism effectively implements sampling. The rest of the function proceeds the same. The pivot now degrades in quality from within $\frac{2n}{9}$ to within $\frac{n}{9}$ of either end of the array. However, a quality check step in QUICKSELECTADAPTIVE turns the sampling mode flag off as soon as one partitioning step did not meet its margin guarantees. Therefore, there may be at most one bad pivot choice per search, and even that is not arbitrarily bad. The time complexity is not affected. This sampling strategy makes QUICKSELECTADAPTIVE adaptive to both $k$ and the data distribution and takes its performance from passable to compelling.

The implementation also contains a few micro-optimizations not shown in the pseudocode: some computations are memoized and sentinel techniques are used in some loops to reduce overheads. The performance gains from most of these optimizations are minor but measurable.

## 7. Experiments and Results

We tested performance on a HP ENVY 810qe Desktop (Intel Core i7 3.6 GHz) machine with 32GB RAM, against arrays of various sizes of 64-bit floating point data (results on 32- and 64-bit integrals are similar). The code and tests were written in the D programming language [1]. The compiler used was the LLVM-based ldc [22] version 0.14.0 (we obtained similar results and rankings with dmd 2.069). We ran each experiment five times (with different random data sets) and took the median of the five timings obtained. The data sets used are:

- Uniformly-distributed random floating-point numbers in the range $-5\,000\,000$ through $5\,000\,000$.
- Numbers generated as above, but sorted prior to testing.
- Normally-distributed data with mean 0 and standard deviation $3\,333\,333$.

Many pivot choosing strategies have been proposed in the literature. We chose the most competitive and in prevalent industrial use, also mentioned in the introduction:

- MEDIAN3: Median of 3 deterministic;
- MEDIAN3RANDOMIZED: Median of 3 randomized;
- NINTHER: Tukey's ninther deterministic; and
- NINTHERRANDOMIZED: Ninther randomized.

The deterministic heuristics are particularly relevant on almost sorted data, where they perform very well. We tried but did not plot two more heuristics: single random pivot performed poorly, and medians of 5 did not perform better than the ninther.

### 7.1 Finding the Median

First we focus on finding the median. Fig. 1 plots the absolute run times of the algorithms tested for finding the median in arrays of uniformly-distributed floating point numbers. To avoid clutter, we did not plot MEDIAN3 and NINTHER because their behavior on random data is similar to their respective randomized versions.

To give a clearer image of the relative performance of the algorithms, Fig. 2 shows the relative speedup of QUICKSELECTADAPTIVE over all other algorithms. Fig. 3 shows speedup only over heuristics-based algorithms.

Ranking is similar on Gaussian-distributed floating point numbers. Fig. 4 shows the speedup of QUICKSELECTADAPTIVE over the others, and Fig. 5 shows its speedup only over heuristics-based algorithms.

One interesting aspect is that REPEATEDSTEPADAPTIVE performed better than REPEATEDSTEP even on this task (searching for the median), where adaptation is not ostensibly involved. The reason is that adaptation still helps, just not in the first iteration: after one or more iterations in QUICKSELECTADAPTIVE's main loop, the index, the array limits, and their relationship changes. For example, if the first partition pass finds a good approximate pivot, the second pass will process a $k$ close to one edge of the array, where adaptive algorithms have an advantage.

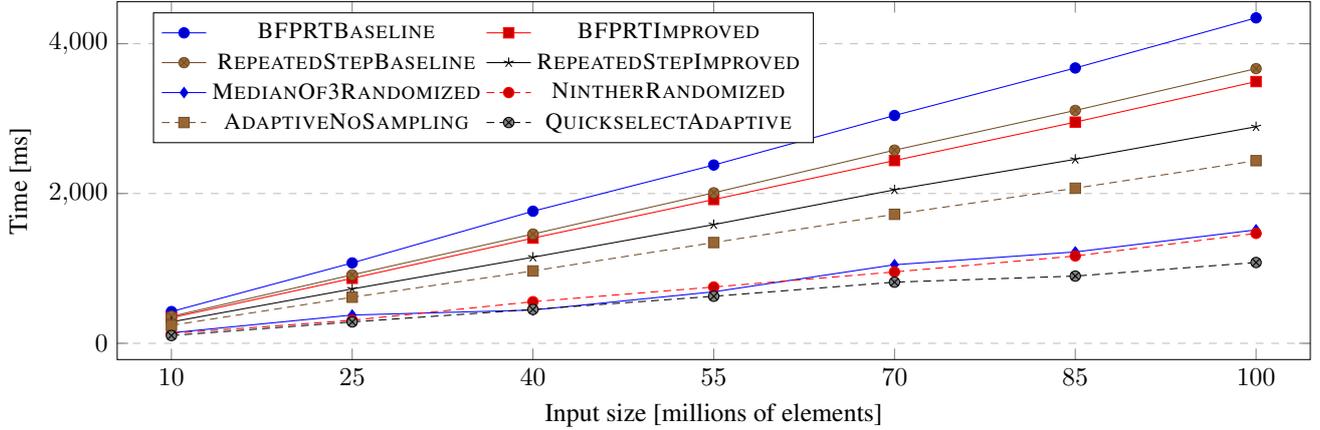

**Figure 1.** Run times of various selection algorithms (64-bit uniformly distributed floating point numbers)

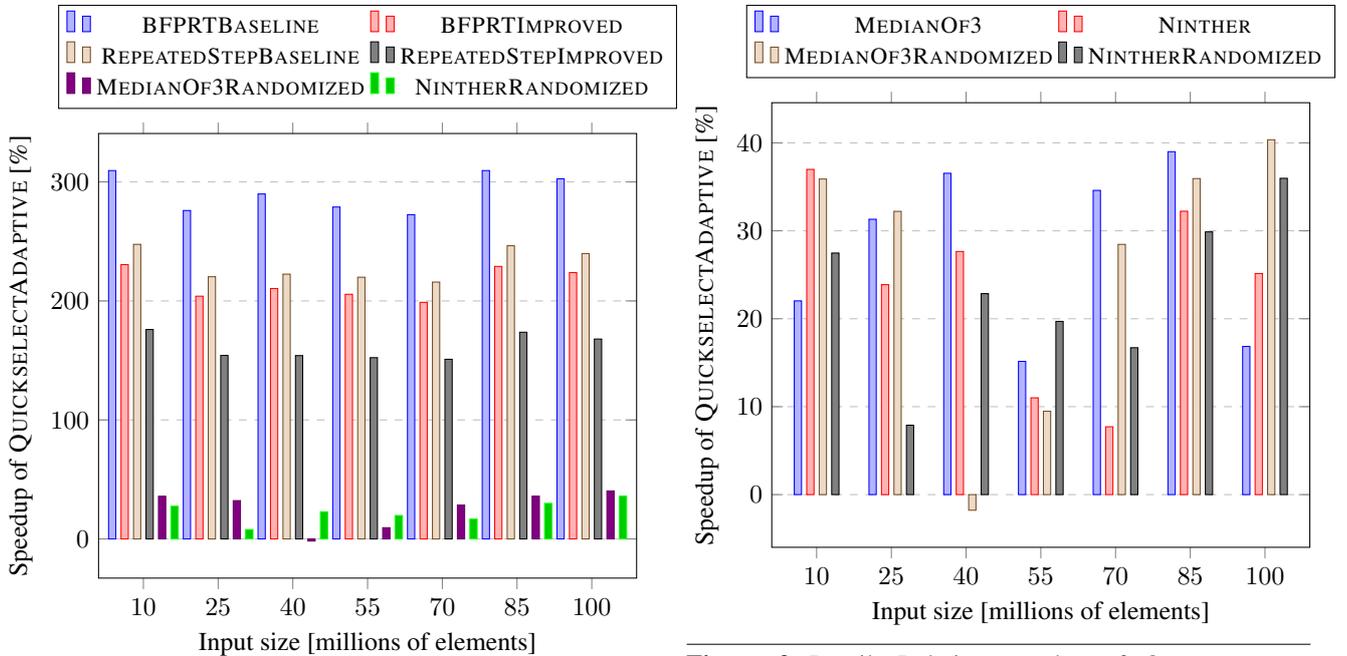

**Figure 2.** Relative speedup of QUICKSELECTADAPTIVE over other algorithms (64-bit uniformly distributed floating point numbers).

**Figure 3.** Detail: Relative speedup of QUICKSELECT-ADAPTIVE over heuristics-based algorithms (64-bit uniformly distributed floating point numbers).

On sorted data, as expected, the deterministic heuristics (MEDIANOF3 and NINTHER) perform the best by a large margin. This is because they are engineered to find the exact median if the data is already sorted, so the cost of QUICKSELECT is essentially that of one pass through HOAREPARTITION. The randomized versions ME-DIANOF3RANDOMIZED and NINTHERRANDOMIZED also perform well on this highly structured input.

### 7.2 Finding Indexes Other than the Median

This experiment measures how adaptation helps in searching for indexes other than the median. This time, $|A|$ is fixed at $1\,000\,000$ and the ratio between $k$ and $|A|$ varies between 5% and 95%. Fig. 8 plots absolute run times for the algorithms tested. Fig. 9 shows the relative speedup of QUICK-SELECTADAPTIVE over the other algorithms. The improvements over non-adaptive MEDIANOFMEDIANS derivatives are strong for all percentiles tested, especially toward the end of the scale, as expected. Fig. 10 gives only the relative speedup over heuristics-based strategies, again showing good improvements. Results on Gaussian-distributed data (not shown) are similar.

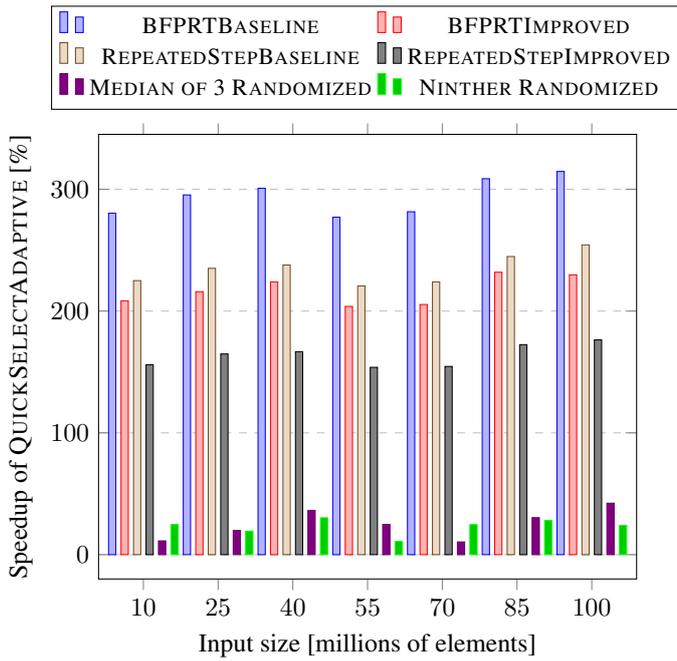

**Figure 4.** Relative speedup of QUICKSELECTADAPTIVE over other algorithms (64-bit Gaussian-distributed floating point numbers).

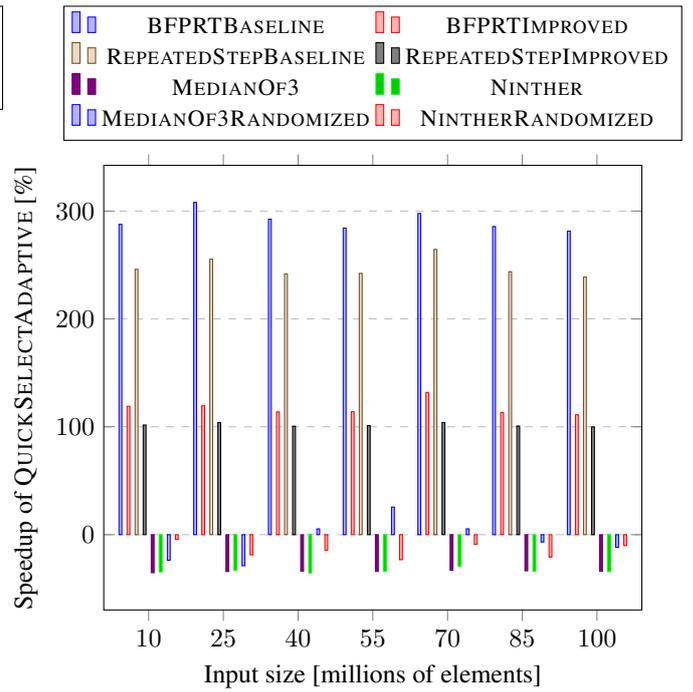

**Figure 6.** Relative speedup (slowdown if negative) of QUICKSELECTADAPTIVE over other algorithms (64-bit sorted floating point numbers).

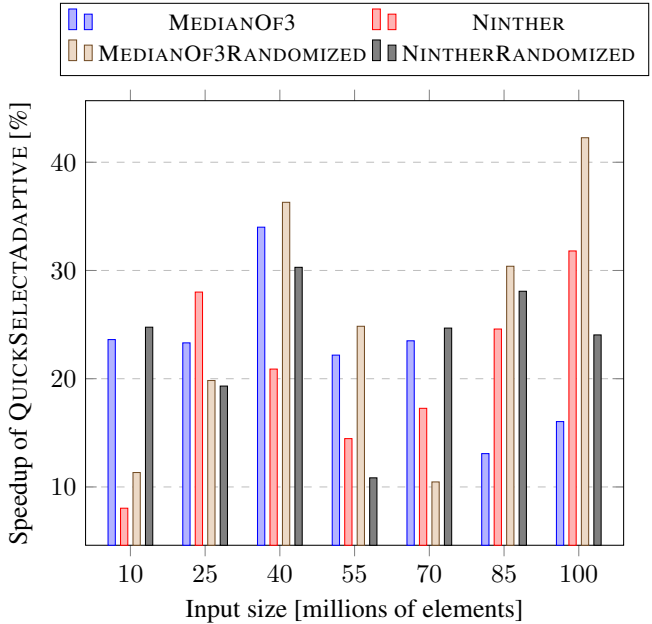

**Figure 5.** Detail: Relative speedup of QUICKSELECTADAPTIVE over heuristics-based algorithms (64-bit Gaussian-distributed floating point numbers).

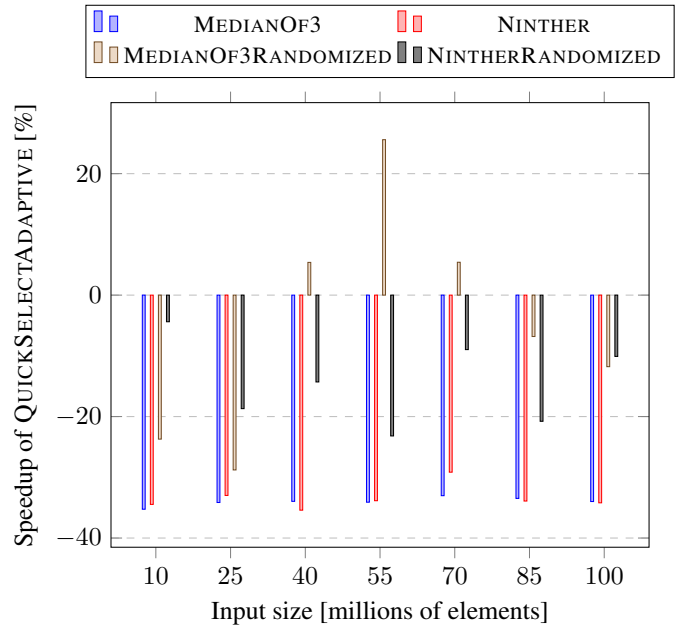

**Figure 7.** Detail: Relative speedup (slowdown if negative) of QUICKSELECTADAPTIVE over heuristics-based algorithms (64-bit sorted floating point numbers).

## 8. Summary and Future Work

We refine the definition of MEDIANOFMEDIANS and its variant REPEATEDSTEP to improve their speed. Measurements show that the best deterministic algorithm QUICKSELECTADAPTIVE outperforms state-of-the-art baselines in the task of partitioning an array around a specified quantile.

There are several future directions we plan to follow. One is to devise principled methods of choosing the group sizes and layouts for each of the specialized algorithms, depending on the index searched. Currently we use three algorithms chosen experimentally, but perhaps a single algorithm with the proper parameterization might replace them all. We also plan to investigate the use of fast deterministic selection algorithms for sorting.

## Acknowledgments

Thanks to Timon Gehr, Ivan Kazmenko, Scott Meyers, Todd Millstein who reviewed drafts of this document. Many participants to the D Programming Languge forum have submitted partition of 5 primitives (http://forum.dlang.org/post/n8u980$1gkr$1@digitalmars.com), of which the one by Teppo Niinimäki was used in this paper.

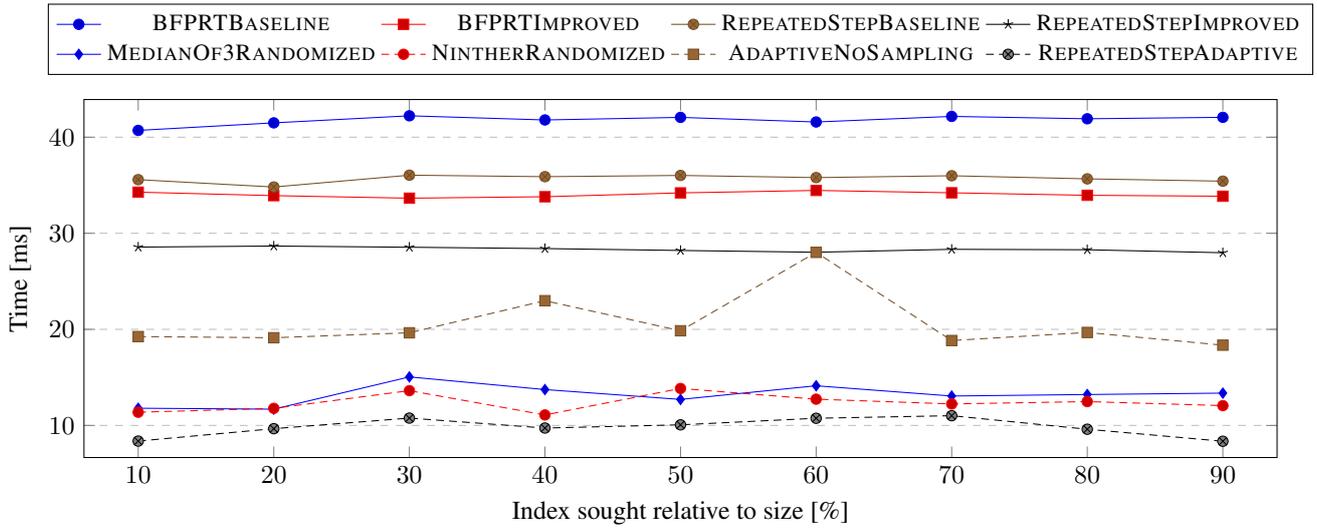

**Figure 8.** Run times of various selection algorithms (64-bit uniformly distributed floating point numbers)

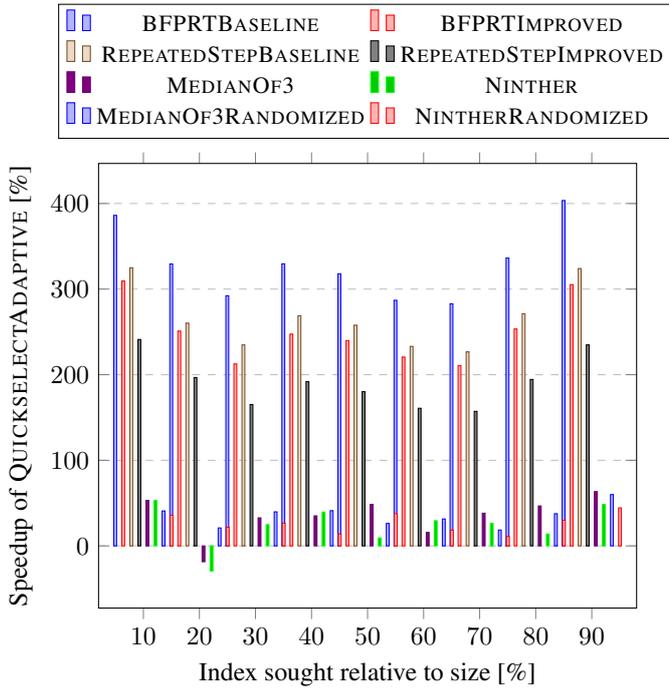

**Figure 9.** Relative speedup of QUICKSELECTADAPTIVE over other algorithms for varying order statistics.

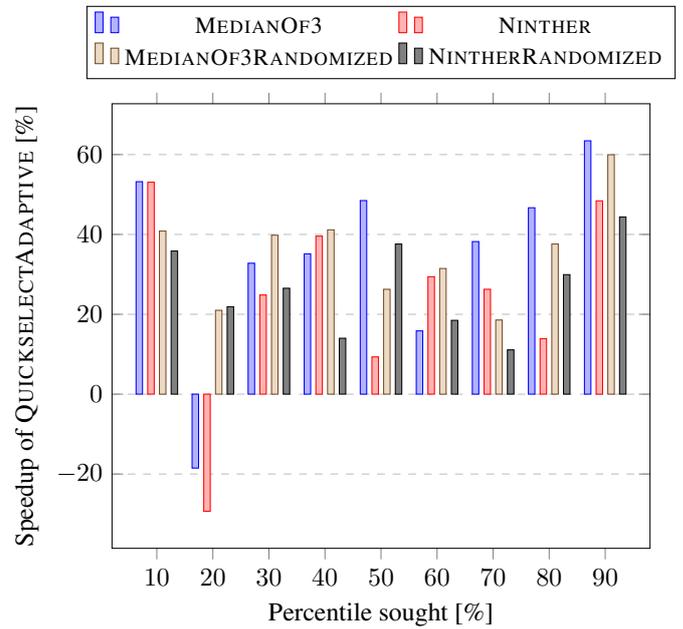

**Figure 10.** Detail: Relative speedup of QUICKSELECT-ADAPTIVE over heuristics-based algorithms for varying order statistics.